%% file: main.tex
\documentclass[aps,prl,reprint,groupedaddress,nofootinbib,longbibliography]{revtex4-2}

\usepackage[T1]{fontenc}
\usepackage{lmodern}
\usepackage{graphicx}
\usepackage{amsmath,amssymb,bm,mathtools}
\usepackage{microtype}
\usepackage{xcolor}
\usepackage{hyperref}
\usepackage{tikz}

\hypersetup{colorlinks=true,citecolor=blue!55!black,linkcolor=blue!55!black,urlcolor=blue!55!black}

\newcommand{\dd}{\mathrm{d}}

\newcommand{\ii}{\mathrm{i}}
\newcommand{\Ord}{\mathcal{O}}

\newcommand{\cD}{\mathcal{D}}

\newcommand{\cL}{\mathcal{L}}

\newcommand{\cR}{\mathcal{R}}

\providecommand{\cA}{\mathcal{A}}
\providecommand{\cF}{\mathcal{F}}

\providecommand{\cV}{\mathcal{V}}

\providecommand{\gE}{\gamma_{\rm E}}

\newcommand{\focussubfigure}[1]{{\setlength{\fboxsep}{2pt}%
\fcolorbox{black!35}{black!5}{#1}}}

\input{diagram_library.tex}
\CausalArrowHaloOff

\begin{document}

\title{Renormalized Dynamical Friction in Black-Hole Spacetimes}

\author{Jordan Wilson-Gerow}
\email[]{jwilsong@andrew.cmu.edu}
\affiliation{Department of Physics, Carnegie Mellon University, Pittsburgh, Pennsylvania 15213, USA}

\date{\today}

\begin{abstract}
We derive the renormalized local equation of motion for the gravitational self-force on a compact object moving through matter in a black-hole spacetime. We integrate out the environment and metric perturbations on a Schwinger--Keldysh contour, obtaining a causal effective action given by a sum of curved-spacetime Feynman diagrams organized by powers of the mass ratio and environmental density. At leading order, an environmental wake gravitationally backreacts on the worldline. The corresponding diagram contains a logarithmic ultraviolet divergence that renormalizes the leading dissipative fluid--particle coupling. We isolate the divergence with a local large-momentum expansion, in which the curved-spacetime retarded Green's function reduces to its flat tangent-space form. We translate the resulting singularity into a Barack--Ori mode-sum scheme, deriving an analytic puncture and regularization parameters for an ideal-fluid environment. Mode subtraction fixes the divergent part but not the finite Wilson coefficient. For collisionless dust, we determine this coefficient by matching to an exact near-zone calculation for a black hole moving through the medium. This gives a complete renormalization prescription suitable for numerical self-force calculations.
\end{abstract}

\maketitle

\section{Introduction}

A compact object moving through matter sources a wake whose gravitational field acts back on the object. In flat spacetime this backreaction is often summarized by a local drag law with a familiar Coulomb logarithm~\cite{Chandrasekhar:1943ys,ostriker1999dynamical}. On a black-hole background, however, the disturbance propagates through both the medium and the curved geometry before returning to the worldline. The resulting backreaction is a causal gravitational self-force modified by the response of the medium. Computing this force therefore requires propagating the wake globally through the curved spacetime and determining its local action on the body.

Existing curved-background calculations address parts of this problem.  Environmental perturbations, wakes, and fluxes on black-hole backgrounds have been calculated in several concrete systems~\cite{BritoShah2023,DysonEtAl2025,DattaMaselli2026}, but these calculations do not propagate the environmental response back to the worldline to obtain the local force. Self-force has  been studied in nonvacuum spacetimes, including electrovac systems and bodies moving through vacuum regions of matter-supported geometries~\cite{ZimmermanPoisson2014, LinzFriedmanWiseman2014,Zimmerman2015,IsoyamaMendesPoisson2016}, but not for a body moving through matter and sourcing a material wake.

To construct the local equation of motion, we build on the vacuum gravitational self-force, derived by Mino, Sasaki, and Tanaka and by Quinn and Wald~\cite{MinoSasakiTanaka1997,QuinnWald1997}.  Barack--Ori mode-sum regularization~\cite{BarackOri2000,Barack2001} and puncture schemes~\cite{BarackGolbournSago2007,PoundWardell2021} made this framework practical for black-hole perturbation theory.  Galley and Hu~\cite{GalleyHu2009}, building on the worldline effective field theory of Goldberger and Rothstein~\cite{GoldbergerRothstein2006}, subsequently gave an EFT derivation of the same vacuum dynamics.  Their formulation shows that, through the leading self-force orders, renormalization of the point-particle theory introduces no new finite Wilson coefficient.  Vacuum self-force divergences therefore require a subtraction prescription, but no further matching to short-distance physics.

The medium changes that conclusion already at leading environmental self-force order. Our generally covariant in-in worldline EFT identified the local dissipative operators that describe direct particle--fluid interactions~\cite{ModrekiladzeRothsteinWilsonGerow2024}. In separate work, we formulated the gravitational problem in flat spacetime by integrating out a dynamical environment and metric perturbations~\cite{Modrekiladze:2026twz}. The leading wake contribution to the gravitational self-force is logarithmically divergent and renormalizes the leading dissipative operator in the fluid--particle EFT. Its finite Wilson coefficient is physical short-distance data. For collisionless dust, we fixed this coefficient by matching to the exact near-zone scattering of a black hole through the medium~\cite{Modrekiladze:2026twz,Traykova:2023qyv}. Thus a subtraction of the singular field is not, by itself, a complete calculation. It specifies a scheme, whereas matching fixes the Wilson coefficient in that scheme. In this problem, EFT is therefore not merely an alternative formulation of the self-force. It shows that an additional worldline operator must be renormalized and identifies the matching calculation required to fix its coefficient.

This Letter extends the environmental in-in construction to black-hole spacetimes and derives a renormalized expression for the leading wake contribution to the local equation of motion. A local momentum expansion proves that the logarithmic coefficient is the flat tangent-space one.  An analytic puncture then translates the EFT counterterm into Barack--Ori mode-sum regularization parameters.  Matching fixes the remaining finite coefficient for a Schwarzschild black hole moving through collisionless dust.

We use $(-,+,+,+)$ signature.  The environment and particle four-velocities are denoted by $u^\mu$ and $v^\mu$, respectively, with
\begin{equation}
 u^2=v^2=-1,
 \qquad
 \gamma=-u\cdot v,
 \qquad
 P(v)^{\mu\nu}=g^{\mu\nu}+v^\mu v^\nu .
 \label{eq:conventions}
\end{equation}
The relative speed is $v_{\rm rel}=\sqrt{1-\gamma^{-2}}$.  We use $q=m/M_\bullet$ for the mass ratio and $\epsilon$ as a bookkeeping parameter for the environmental stress-energy, $T^{\mu\nu}_{\rm env}=\mathcal{O}(\epsilon)$.

\section{Covariant in-in EFT}

We begin from the conservative action
\begin{equation}
 S[g,z,e,\Psi]=S_{\rm EH}[g]+S_{\rm pp}[g,z,e]+S_{\rm env}[\Psi,g]+S_{\rm wl}^{\rm fs},
 \label{eq:microscopic-action}
\end{equation}
with the point particle written in first-order (einbein) form,
\begin{equation}
 S_{\rm pp}=\frac12\int\dd\lambda
 \left[e^{-1}g_{\mu\nu}\dot z^\mu\dot z^\nu-e\,m^2\right].
 \label{eq:polyakov}
\end{equation}
We use $e=1/m$, for which the einbein equation gives $g_{\mu\nu}\dot z^\mu\dot z^\nu=-1$.  The conservative finite-size operators in $S_{\rm wl}^{\rm fs}$ enter beyond the order considered here and will be omitted.

Doubling the fields and histories and integrating out the environment in its initial state $\rho_{\rm env}$ gives its contribution $W[g_1,g_2]$ to the in-in effective action for the metric,
\begin{equation}
 e^{\ii W[g_1,g_2]}
 =\int\cD\Psi_1\cD\Psi_2\,\rho_{\rm env}
 e^{\ii S_{\rm env}[\Psi_1,g_1]-\ii S_{\rm env}[\Psi_2,g_2]} .
 \label{eq:influence-functional}
\end{equation}
We decompose $g_{i}=\bar{g}+h_i$, where $\bar{g}_{\mu\nu}$ is a vacuum black-hole solution, $G_{\mu\nu}[\bar{g}]=0$. The mean environmental correction to this geometry is treated perturbatively through the one-point vertex $W_1\equiv\delta W/\delta g\big|_{\bar g}$, with tensor and Schwinger--Keldysh labels suppressed. The doubled EFT must also contain the local operators allowed by covariance and the power counting. Dissipative particle--environment operators do not factorize into a difference of two conservative actions and are therefore not contained in Eq.~\eqref{eq:microscopic-action}. Following Ref.~\cite{ModrekiladzeRothsteinWilsonGerow2024}, we denote the leading such contribution by $S^{\rm in\text{-}in}_{\rm loc}[z_1,z_2;\bar{g},u]$. At the order considered here it is independent of $h_i$, so integrating out the metric perturbations gives
\begin{align}
e^{\ii\Gamma[z_1,z_2]}
={}&e^{\ii S_{\rm loc}^{\rm in\text{-}in}[z_1,z_2;\bar g,u]}\nonumber\\
\times&
\int\cD h_1\cD h_2\,
\exp\big[\ii S_{\rm EH}[\bar g+h_1]-\ii S_{\rm EH}[\bar g+h_2]
\nonumber\\[-2pt]
&+\ii S_{\rm pp}[\bar g+h_1,z_1,e_1]
-\ii S_{\rm pp}[\bar g+h_2,z_2,e_2] \nonumber \\
&+\ii W[\bar g+h_1,\bar g+h_2]\big].
\label{eq:worldline-effective-action}
\end{align}
Gauge-fixing and ghost factors are implicit.  We write $z_r^\mu$ for the physical history and $z_a^\mu$ for the covariant difference of the two histories, with the precise construction given in Refs.~\cite{GalleyTiglio2009,Galley2013,ModrekiladzeRothsteinWilsonGerow2024}.  The physical limit is $z_{a}=0$, and the equation of motion follows from $\delta\Gamma/\delta z_a^\mu|_{z_a=0}=0$. Its causality follows from the closed-time-path contour.

The effective action $\Gamma$ is the sum of connected in-in Feynman diagrams.  In the expansion of $W$ about $\bar g_{\mu\nu}$, the one-point vertex $W_1\sim\overline T^{\mu\nu}$ describes the mean environmental stress tensor, while $W_2\sim\Pi_{\rm ret}$ describes its retarded linear response. Higher variations $W_{n\geq3}$ describe nonlinear environmental response and begin contributing to the effective action at $\Ord(\epsilon q^3)$~\cite{Modrekiladze:2026drr}.

Because the point-particle graviton vertex is linear in the metric, the sum of metric-mediated diagrams ending on the worldline has the form of minimal coupling to an effective metric. Before reparameterization, the equation of motion they generate is therefore a geodesic equation in $g^{\rm eff}_{\mu\nu}=\bar g_{\mu\nu}+h^{\rm ret}_{\mu\nu}$.  Writing the effective connection in terms of the background connection and parameterizing the trajectory by background proper time produces the projector $P(v)^{\mu\nu}$,
\begin{align}
 m\bar\nabla_v v^\mu&=F^\mu_{\rm met}[h^{\rm ret}]+F^\mu_{\rm loc},
 \nonumber\\
 F^\mu_{\rm met}[h^{\rm ret}]&=-\frac{m}{2}P(v)^{\mu\nu}
 \left(2\bar\nabla_\alpha h^{\rm ret}_{\nu\beta}
       -\bar\nabla_\nu h^{\rm ret}_{\alpha\beta}\right)
 v^\alpha v^\beta .
 \label{eq:renormalized-EOM}
\end{align}
Here $F^\mu_{\rm loc}$ arises directly from local in-in worldline operators rather than from graviton exchange. For later use, we write $F_{\rm met}^{\mu}[h]\equiv\cF^{\mu A}_{x}h_{A}(x)\big|_{x=z}$, where $A=(\alpha\beta)$ denotes a symmetric tensor index pair. The metric-mediated force is a functional of the past history and is therefore nonlocal in time.  At $\epsilon=0$ it reduces to the vacuum in-in self-force expansion~\cite{GalleyHu2009}.

\begin{figure}[t]
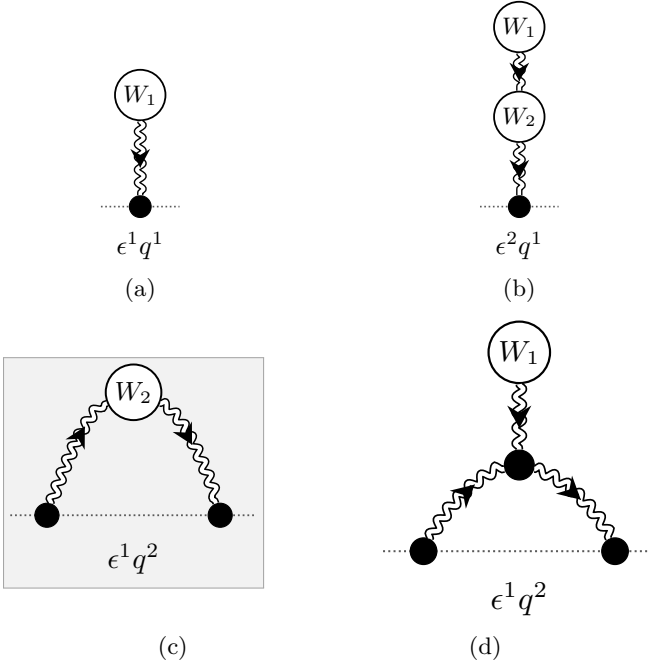

\centering
\begin{minipage}[b]{0.42\columnwidth}\centering
\FNineDOneSKMeanMetric
\par\smallskip (a)\end{minipage}\hfill
\begin{minipage}[b]{0.42\columnwidth}\centering
\FNineDTwoSKResponseChain
\par\smallskip (b)\end{minipage}
\par\medskip
\focussubfigure{\resizebox{0.38\columnwidth}{!}{\FTenDOneSKEnvironmentalResponse}}\hfill
\resizebox{0.42\columnwidth}{!}{\FTenDTwoSKMeanMetricNonlinearOneSF}
\par\smallskip
(c)\hspace{0.43\columnwidth}(d)
\caption{Environmental in-in diagrams in the $(\epsilon,q)$ expansion: (a) the mean metric at $\Ord(\epsilon q)$, (b) its response at $\Ord(\epsilon^2q)$, and the $\Ord(\epsilon q^2)$ contributions from (c) polarization through $W_2$ and (d) coupling to $W_1$ through a cubic graviton vertex.  Dotted lines denote worldlines, doubled wavy lines denote curved-space gravitons, and arrowheads indicate causal Schwinger--Keldysh components.  We renormalize the shaded topology.}
\label{fig:environmental-expansion}
\end{figure}

Figure~\ref{fig:environmental-expansion} organizes the environmental diagrams by effective-action order.  At $\Ord(\epsilon q)$, the particle couples to the mean metric sourced by $W_1$, as in the background effects studied in Ref.~\cite{DattaMaselli2026}.  The upper row advances in $\epsilon$ at fixed $q$, while the lower row contains the two $\Ord(\epsilon q^2)$ force topologies.

\begin{figure}[t]
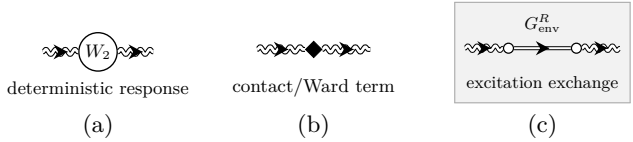

\centering
\begin{minipage}[b]{0.30\columnwidth}\centering
\scalebox{0.76}{\DFDOneWTwoResponse}
\par\smallskip (a)\end{minipage}\hfill
\begin{minipage}[b]{0.30\columnwidth}\centering
\scalebox{0.78}{\DFDTwoWTwoContact}
\par\smallskip (b)\end{minipage}\hfill
\begin{minipage}[b]{0.34\columnwidth}\centering
\focussubfigure{\scalebox{0.72}{\DFDThreeWTwoExchange}}
\par\smallskip (c)\end{minipage}
\caption{The response $W_2$ in (a) separates into (b) local contact terms and (c) exchange through $G^R_{\rm env}$.  We renormalize the shaded exchange term; only (b) plus (c) is gauge complete.}
\label{fig:W2-split}
\end{figure}

In panel (c), the environment polarizes the particle's retarded perturbation.  In panel (d), that perturbation couples to the mean environmental geometry through a cubic graviton vertex.  Our ultraviolet calculation concerns panel (c).  The induced metric for this topology has the schematic form
\begin{equation}
 h^{\rm wake}_{\epsilon q}
 =(16\pi G)^2G_{\rm ret}\circ\Pi_{\rm ret}\circ G_{\rm ret}\circ T_{\rm pp} .
 \label{eq:polarization-field}
\end{equation}
The full environmental response $\Pi_{\rm ret}$ contains excitation-exchange as well as contact terms from varying the stress tensor and volume element~\cite{Kovtun2012}, as shown in Fig.~\ref{fig:W2-split}.  Only their sum obeys the linearized diffeomorphism Ward identity.  We evaluate the exchange contribution.  The contact and cubic-graviton topologies have the same superficial ultraviolet power counting.  In both cases the leading large-momentum term is odd under reversal of the local momentum and vanishes under symmetric integration.  The first nonzero parity-even term is ultraviolet convergent, so neither contribution changes the logarithmic coefficient $C^\mu$ derived below.

Amputating the return graviton in Fig.~\ref{fig:environmental-expansion}(c) leaves $\Pi_{\rm ret}\circ G_{\rm ret}\circ T_{\rm pp}$, the particle-driven environmental stress tensor shown in Fig.~\ref{fig:env-stress}(c).  This is the wake computed in Refs.~\cite{BritoShah2023,DysonEtAl2025}.  Reattaching the return propagator and applying the linearized force operator $\cF_x^{\mu A}$ defined by Eq.~\eqref{eq:renormalized-EOM} returns that wake to the worldline.  This is an amputation rather than a cut.  All remaining propagators are retarded, and no line is placed on shell.

\begin{figure}[t]
\centering
\begin{minipage}[b]{0.22\columnwidth}\centering
\scalebox{0.80}{\FFourDOneSKStressMean}
\par\smallskip (a)\end{minipage}\hfill
\begin{minipage}[b]{0.36\columnwidth}\centering
\scalebox{0.80}{\FFourDTwoSKStressSelfGravity}
\par\smallskip (b)\end{minipage}\hfill
\begin{minipage}[b]{0.36\columnwidth}\centering
\scalebox{0.80}{\FFourDThreeSKStressDrivenByParticle}
\par\smallskip (c)\end{minipage}
\caption{Environmental stress tensor $\langle T^{\mu\nu}_{\rm env}\rangle$, conventions as in Fig.~\ref{fig:environmental-expansion}: (a) the equilibrium configuration, (b) its self-gravitating correction, and (c) the particle-driven wake at $\Ord(\epsilon q)$.}
\label{fig:env-stress}
\end{figure}

By contrast, cutting $\Pi_{\rm ret}$ on its spectral function identifies energy transferred to the medium.  Energy stored in the wake need not reach the horizon or null infinity, so flux balance constrains the orbit-averaged absorptive force only when the energy deposited into and stored in the environment is included.  It does not determine the reactive force or the instantaneous local field.

\section{Local ultraviolet region}

The wake metric in Eq.~\eqref{eq:polarization-field} contributes a force $F^{\mu}_{\rm wake}\equiv F^{\mu}_{\rm met}[h^{\rm wake}_{\epsilon q}]$. Restricting to the shaded exchange diagram and evaluating the force at $z_0 = z(\tau_0)$, we obtain
\begin{equation}
\begin{aligned}
 F^\mu_{\rm wake}(\tau_0)
 ={}&\int \dd\tau'\dd V_y\dd V_{y'}\,
 \cF_x^{\mu A}G^{\rm ret}_{AB}(x,y)
 \\
 &\quad\times\Pi_{\rm ret}^{BC}(y,y')
 G^{\rm ret}_{CD}(y',z')J^D(\tau')
 \bigg|_{x=z_0}\,.
\end{aligned}
 \label{eq:formal-force}
\end{equation}
Here $J^D=(m/2)v^\rho v^\sigma$, with $D=(\rho\sigma)$, is the point-particle graviton source in the conventions of Eq.~\eqref{eq:polyakov}. To isolate the ultraviolet divergence, we split the integration region with a smooth window supported in a convex normal neighborhood of $z_0$.  The complement is ultraviolet finite and retains the global propagation, environmental profile, and orbital history.

In the local region we use Riemann normal coordinates (RNC), $Y^a$, centered at $z_0$. We Fourier transform the component kernels in their coordinate separations and expand the slowly varying coefficients about $z_0$.  The Lorenz-gauge retarded graviton then has the local large-momentum expansion
\begin{equation}
\begin{aligned}
 G^{\rm ret}_{ab,cd}(Y_1,Y_2)
 ={}&\int_k e^{\ii k\cdot(Y_1-Y_2)}
 \bigg[
 \frac{\mathbb P_{ab,cd}}{k^2+\ii0\,k\cdot t}
 \\
 &\quad+\Ord\!\left(\frac{R}{k^4}\right)
 +\Ord\!\left(\frac{\bar\nabla R}{k^5}\right)
 \bigg],
\end{aligned}
 \label{eq:local-RNC-graviton}
\end{equation}
with all coefficient tensors evaluated at $z_0$.  This is a local asymptotic representation, not a global Fourier transform of the black-hole Green function.

At leading order the RNC metric is $\eta_{ab}$, the volume element is flat, and the local kernels depend only on coordinate differences.  Substitution in Eq.~\eqref{eq:formal-force} therefore makes the $y$ and $y'$ integrations set the local momenta equal.  Replacing the worldline by its tangent line gives the near-coincidence contribution to the force
\begin{equation}
\begin{aligned}
 F^\mu_{\rm near}
 ={}&\int_k 2\pi\delta(k\cdot v)\,
 \cV^{\mu A}_0(k)G^{\rm ret}_{0,AB}(k)
 \\
 &\quad\times\Pi^{BC}_{\rm ret,0}(k)
 G^{\rm ret}_{0,CD}(k)J_0^D
 \\
 &\quad\times\left[1+\Ord\!\left(\frac{1}{k\cL}\right)
 +\Ord\!\left(\frac{1}{(k\cR)^2}\right)\right].
\end{aligned}
 \label{eq:local-momentum-force}
\end{equation}
The leading integrand is the flat tangent-space expression evaluated on the local state at $z_0$. Gradients of the environment over a scale $\cL$ give the first correction. Corrections due to the curved geometry enter at relative order $R/k^2\sim(k\cR)^{-2}$, where $R$ is shorthand for the Riemann tensor and $\cR$ is the typical curvature scale~\cite{BunchParker1979,Bunch1981,BuchbinderEtAl1984,GalleyHu2009,Zimmerman2015}.  The momentum delta functions express an effective translation invariance in the local tangent space, not a global spacetime symmetry.

Here $\cV_0^{\mu A}(k)$ is the flat tangent-space momentum-space force vertex obtained by applying $\cF_x^{\mu A}$ to the return graviton, and is linear in $k$.  The tangent worldline makes the sourced perturbation stationary in the particle's local rest frame and produces $2\pi\delta(k\cdot v)$.  The absorptive fluid response is supported on the sound cone, $\omega^2=c_s^2|\bm k|^2$.  Imposing these conditions leaves a marginal radial integral~\cite{Modrekiladze:2026twz}:
\begin{equation}
 \dd F_0^\mu=C^\mu\frac{\dd k}{k},
 \qquad
 \dd F_{\nabla}^\mu\sim\frac{\dd k}{k^2\cL},
 \qquad
 \dd F_R^\mu\sim \frac{\dd k}{k^3\cR^2 } .
 \label{eq:uv-hierarchy}
\end{equation}
Only the first term is UV divergent.  Thus the coefficient of the curved-space logarithm is precisely the flat-space coefficient evaluated on the local state.  This conclusion relies on the fact that the flat diagram is already just marginally divergent.  Curvature-dependent logarithms can occur in other diagrams with a larger superficial degree of divergence.  We find no such diagrams at the order considered here.

For an ideal fluid, Ref.~\cite{Modrekiladze:2026twz} gives
\begin{align}
 C^\mu&=\cA(\gamma,v_{\rm rel},c_s)\cD^\mu,
 \label{eq:C-general}\\
 \cD^\mu&=P(v)^\mu{}_{\nu}u^\nu=u^\mu-\gamma v^\mu,
 \nonumber\\
 \cA&=4\pi G^2(\rho_E+p_E)m^2
 \frac{(2\gamma^2-1)^2}{(\gamma^2-1)\gamma v_{\rm rel}}
 \Theta(v_{\rm rel}-c_s) .
 \label{eq:A-general}
\end{align}
$\cD^\mu$ gives the drag direction.  We retain general $\rho_E$, $p_E$, and $c_s$ here and specialize to collisionless dust only for the finite matching below.

The logarithmic ultraviolet divergence in Eq.~\eqref{eq:uv-hierarchy} renormalizes a local worldline operator. The corresponding bare action and force are
\begin{equation}
 S_K=\int\dd\tau\,K_{\rm bare}(\gamma)\,u_\mu X_a^\mu,
 \qquad
 F_K^\mu=\cD^\mu K_{\rm bare}(\gamma)\,.
 \label{eq:K-operator}
\end{equation}
Here $X_a^\mu=P(v)^\mu{}_{\nu}z_a^\nu$, as required by reparameterization invariance. This is the leading local dissipative particle--fluid operator in the generally covariant in-in worldline EFT of Ref.~\cite{ModrekiladzeRothsteinWilsonGerow2024}.  In~\cite{Modrekiladze:2026twz} we showed that the flat-spacetime gravitational wake renormalizes its Wilson coefficient.  Here the same operator absorbs the local divergence of the curved-spacetime diagram.  The EFT force is the sum of the loop diagram and this bare-operator contribution.

\section{Mode-sum renormalization}

We now translate the local momentum-space counterterm into the angular mode-sum used in Schwarzschild perturbation theory. We specialize to a circular Schwarzschild orbit at areal radius $r_0$. For the local angular analysis we rotate the sphere so that the particle lies at the pole and write $\Omega=(\alpha,\beta)$, where $\alpha$ is the angular separation from the particle and $\beta$ is the azimuth about it.  Extending a regular tetrad off the worldline, define
\begin{align}
 \mathfrak F^{\hat a}(t_0,r,\Omega)
 &=e^{\hat a}{}_{\mu}(x)\cF_x^{\mu A}h_A^{\rm wake}(x),
 \label{eq:extended-force-component}\\
 F_{\ell m}^{\hat a}(r)
 &=\int\dd\Omega\,Y_{\ell m}^*(\Omega)\mathfrak F^{\hat a}(t_0,r,\Omega),
 \nonumber\\
 F_\ell^{\hat a}(r)
 &=\sum_{m=-\ell}^{\ell}F_{\ell m}^{\hat a}(r)Y_{\ell m}(\Omega_p),
 \qquad L=\ell+\tfrac12 .
 \label{eq:force-mode-definition}
\end{align}
The Appendix derives the local point-split field and projects it onto the modes in Eq.~\eqref{eq:force-mode-definition}.  The essential point is that, at large $L$, the $m$-summed Legendre projection probes angular separations $\alpha\sim L^{-1}$, where the local tangent-space approximation in Eq.~\eqref{eq:local-momentum-force} applies, and therefore transverse momenta are of order $\kappa\sim L/r_0$. Since the ultraviolet measure is $\dd\kappa/\kappa$, its angular projection gives
\begin{equation}
 F_\ell^{\hat a}=\frac{C^{\hat a}}{L}+\Ord(L^{-2}) .
 \label{eq:Fell-result}
\end{equation}
The coefficient $C^{\hat a}$ is the Barack--Ori regularization parameter associated with the new in-medium singularity. 

The sum of Eq.~\eqref{eq:Fell-result} therefore contains the divergence $C^{\hat a}\ln N$.  We split the bare coefficient and total renormalized force as
\begin{align}
 K_{\rm bare}(\gamma;N)=&\sum_{\ell=0}^N K_{{\rm ct},\ell}+K_{\rm ren}(r_0),
 \label{eq:bare-split}\\
 F^{\hat a}_{\rm ren}
 =\lim_{N\to\infty}&\sum_{\ell=0}^{N}
 \left(F_\ell^{\hat a}+\cD^{\hat a}K_{{\rm ct},\ell}\right)
 +\cD^{\hat a}K_{\rm ren}(r_0) .
 \label{eq:bare-force-split}
\end{align}
The $\gamma$ dependence is suppressed in $K_{{\rm ct},\ell}$ and $K_{\rm ren}$. The mode sum in Eq.~\eqref{eq:bare-force-split} is the loop-diagram contribution after counterterm subtraction, while the final term is the force generated by the finite renormalized coefficient.  The subtraction fixes the divergent part of $K_{\rm bare}$ but cannot determine $K_{\rm ren}$.  To make this prescription usable mode by mode, we choose the rotationally covariant logarithmic puncture
\begin{align}
 K_{\rm ct}(\alpha)&=\frac{\cA}{2}\ln(1-\cos\alpha),
 \nonumber\\[-2pt]
 K_{{\rm ct},\ell}&=
 \begin{cases}
 \displaystyle -\cA\,\frac{L}{\ell(\ell+1)},&\ell\geq1,\\[3pt]
 \displaystyle \frac{\cA}{2}(\ln2-1),&\ell=0.
 \end{cases}
 \label{eq:K-ct-puncture}
\end{align}
The sequence $K_{{\rm ct},\ell}$ is defined at the worldline, while $K_{\rm ct}(\alpha)$ extends it over the sphere.  We add $F^{\hat a}_{\rm punc}=\cD^{\hat a}K_{\rm ct}$ to the extended force before performing the mode sum.\footnote{Add vs. subtract is just a convention difference between EFT and GSF.}  Other off-worldline extensions change only a finite remainder that is absorbed into $K_{\rm ren}$.  The Appendix shows explicitly that this puncture subtraction is the Legendre representation of the worldline counterterm.  In this sense, mode-sum regularization and EFT renormalization are the same subtraction written in two different bases.

\section{Flat-space matching}
\label{sec:matching}

The $1/L$ tail in Eq.~\eqref{eq:Fell-result} is the local logarithmic singularity in the off-worldline extension of the wake force, $\mathfrak{F}^{\hat a}(x)$, represented in angular-harmonic space. The Appendix derives this singularity and shows that the renormalized mode sum can be evaluated by adding the puncture to the point-split force before taking the coincidence limit:
\begin{align}\label{eq:collapsed-sum}
\sum_{\ell=0}^{\infty}
\left(F_\ell^{\hat a}
+\cD^{\hat a}K_{{\rm ct},\ell}\right)
&=
\lim_{\alpha\to0}
\left[
\bar{\mathfrak F}^{\hat a}(\alpha)
+F_{\rm punc}^{\hat a}(\alpha)
\right] \\
&= C^{\hat a}\left[ \ln\frac{\Lambda}{r_0} +\ln\frac{\sqrt2}{1+s} \right]\,,
\end{align} 
where $\Lambda\equiv2e^{-\gE}/k_{\min}$ is the infrared length associated with the hard lower momentum cutoff $k_{\min}$, and $s=c_s/v_{\rm rel}$. Here \(\bar{\mathfrak F}^{\hat a}(\alpha)\) is the azimuthal average of $\mathfrak{F}^{\hat a}(x)$.  While the puncture/counterterm removes the logarithmic singularity, the finite Wilson coefficient $K_{\rm ren}(r_0)$ remains to be determined. We fix it by matching to near-zone dynamics.

Any exact near-zone result for the same straight-line problem, computed to all orders in $G$ and regulated in the same infrared scheme, can be written as
\begin{equation}
 F^{\mu}_{\rm exact}
 =C^\mu\left[\ln\frac{\Lambda}{\ell_{\rm UV}(v_{\rm rel},c_s)}
 +\cR(v_{\rm rel},c_s)\right],
 \label{eq:schematic-matching}
\end{equation}
where $\ell_{\rm UV}$ is a physical short-distance scale and $\cR$ is the finite remainder in this infrared scheme.  Equating the two forces cancels $\Lambda$ and gives
\begin{align}
 K_{\rm ren}(r_0)=\cA\bigg[\ln\frac{r_0}{\ell_{\rm UV}(v_{\rm rel},c_s)}
 -\ln\frac{\sqrt2}{1+s}+\cR(v_{\rm rel},c_s)\bigg]\,.
 \label{eq:Kren-general-s}
\end{align}
The $r_0$ dependence of Eq.~\eqref{eq:Kren-general-s} is precisely the renormalization-group running required to cancel the fiducial scale in Eq.~\eqref{eq:collapsed-sum}---the straight-line force cannot depend on the choice of $r_0$.  The remaining terms are the finite conversion between the angular puncture scheme and the physical near-zone result.

No exact result of this form is presently available for $c_s>0$.  For collisionless dust, Ref.~\cite{Traykova:2023qyv} gives the required result from geodesic scattering, regulated by a hard upper cutoff $b_{\rm max}$ on the transverse impact parameter. In Ref.~\cite{Modrekiladze:2026twz} we matched this result to the flat-spacetime EFT and extracted the finite remainder.

The dust response sets $k_\parallel=0$, so $k_{\min}$ is a transverse-momentum cutoff. Comparing the two sharp schemes gives $k_{\min}=2e^{-\gE}/b_{\rm max}$, as derived in the Appendix, and hence $\Lambda=b_{\rm max}$.  In this case $\ell_{\rm UV}=R_c\equiv Gm/v_{\rm rel}^2$ is the impact parameter for a Newtonian $\pi/2$ deflection and $\cR_{\rm match}(v_{\rm rel})$ is the fitted finite remainder.  We then obtain
\begin{equation}
 K_{\rm ren}(r_0)=\cA\left[\ln\frac{r_0}{R_c}-\frac12\ln2
 +\cR_{\rm match}(v_{\rm rel})\right]\qquad(\text{dust}).
 \label{eq:Kren-shift}
\end{equation}
Equations~\eqref{eq:bare-force-split} and \eqref{eq:Kren-shift} therefore give the renormalized force for collisionless dust.  The numerically computed black-hole modes enter the subtracted sum in Eq.~\eqref{eq:bare-force-split}, while Eq.~\eqref{eq:Kren-shift} supplies its matched finite worldline contribution.

\section{Discussion}

We have derived and renormalized the local equation of motion describing the wake-mediated gravitational self-force on a compact object moving through a fluid in a black-hole spacetime. The expansion of the in-in effective action identifies the diagram in which the particle-sourced perturbation propagates through the medium and returns to the worldline.  Its ultraviolet region is controlled by a flat tangent-space integrand, while the remaining mode calculation retains the black-hole geometry, boundary conditions, environmental profile, and orbital history.  The logarithmic UV divergence appears as a $1/L$ tail in the angular modes, and we compute the corresponding Barack--Ori regularization parameter analytically. This divergence is removed by a puncture that implements the EFT counterterm, and the remaining finite Wilson coefficient is fixed by near-zone matching.

This is where EFT changes the calculation relative to the leading vacuum self-force.  In vacuum, the worldline theory reproduces the MiSaTaQuWa dynamics, but effacement implies that no terms beyond minimal coupling are renormalized through the leading self-force orders~\cite{GalleyHu2009}.  Here, by contrast, the leading in-medium diagram already renormalizes the dissipative operator in Eq.~\eqref{eq:K-operator}.  Barack--Ori subtraction fixes its singular large-$\ell$ tail, but it cannot determine the finite Wilson coefficient multiplying that operator.  A subtraction without matching would merely choose a renormalization condition, leaving the short-distance physics unresolved.

The tail in Eq.~\eqref{eq:Fell-result}, the puncture in Eq.~\eqref{eq:K-ct-puncture}, and the matching relation in Eq.~\eqref{eq:Kren-general-s} apply to an ideal fluid whenever $v_{\rm rel}>c_s$.  Collisionless dust is needed only to supply the presently available exact near-zone matching calculation.  For $c_s>0$, the same prescription applies once the corresponding physical matching result is available.

Several pieces of the complete $\Ord(\epsilon q^2)$ force remain outside the present calculation. These are the contact part of $\Pi_{\rm ret}$, the mean-metric topology of Fig.~\ref{fig:environmental-expansion}(d), and the accretion sector.  The coefficient of the logarithmic divergence computed here is not affected by these contributions, although they are required for a complete force calculation.  Kerr, eccentric motion, and more general distribution functions change the global implementation, not the local power counting that fixes the ultraviolet logarithm.

To our knowledge, this is the first derivation from first principles of the causal local equation of motion describing the in-medium gravitational self-force in a black-hole spacetime, together with a local regularization and matching prescription suitable for numerical evaluation.  It is the local-force counterpart of recent curved-background wake and flux calculations~\cite{BritoShah2023,DysonEtAl2025,DattaMaselli2026}: the environmental perturbation is propagated back to the worldline and renormalized to yield a gravitational self-force.

\textit{Note added.---} While this manuscript was in preparation,
Ref.~\cite{Datta2026EnvironmentalGSF} appeared, emphasizing the interpretation of dynamical friction as an environmental gravitational self-force in flat-spacetime perturbation theory.  The present work instead focuses on the derivation and renormalization of this local equation of motion in a black-hole spacetime.

\section{Acknowledgments}
We thank Ira Rothstein and Beka Modrekiladze for extensive discussions and collaboration on related work. The author is supported by the U.S. Department of Energy grant DE-SC001011, and by a President's Postdoctoral Fellowship at CMU.

\bibliographystyle{apsrev4-2}
\bibliography{references}

\clearpage
\appendix
\section{APPENDIX: Angular tail and mode-scheme matching}

\subsection{Large-\texorpdfstring{$\ell$}{l} angular tail}

By the addition theorem, the $m$-summed mode in Eq.~\eqref{eq:force-mode-definition} is
\begin{equation}
 F^{\hat a}_\ell(r)=\int\dd\Omega\,\frac{2\ell+1}{4\pi}P_\ell(\cos\alpha)\,
 \mathfrak F^{\hat a}(t_0,r,\Omega),
 \label{eq:addition-theorem}
\end{equation}
where $\alpha$ is the angular separation from the particle.  We take the tetrad extension to be constant on the tangent space across the RNC region.  Since $e^{\hat a}{}_{\mu}(z_0)=\delta^{\hat a}{}_{\mu}$, this gives $C^{\hat a}=C^\mu$ at $z_0$, while $\hat k$ and $\Delta\bm x$ below are ordinary flat three-vectors.

Undoing the coincidence limit in Eq.~\eqref{eq:uv-hierarchy} gives the local point-split field [cf. Eq.~(6.3) of Ref.~\cite{Modrekiladze:2026twz}],
\begin{equation}
 \mathfrak F^{\hat a}(x)
 =C^{\hat a}\,\Re\!\int_{k_{\min}}^{\infty}\frac{\dd\kappa}{\kappa}
 \int_0^{2\pi}\frac{\dd\psi}{2\pi}\;e^{\ii\bm k\cdot\Delta\bm x} .
 \label{eq:F-offworldline}
\end{equation}
The support conditions restrict $\hat k$ to
\begin{equation}
 \hat k=s\,\hat e_{\hat\phi}
 +\sqrt{1-s^2}\left(\cos\psi\,\hat e_{\hat r}+\sin\psi\,\hat e_{\hat\theta}\right),
 \qquad s=\frac{c_s}{v_{\rm rel}},
 \label{eq:cone-parametrization}
\end{equation}
while a field point at radial offset $\delta$ and angular position $(\alpha,\beta)$ has
\begin{equation}
 \Delta\bm x=\delta\,\hat e_{\hat r}
 +r_0\alpha\left(\cos\beta\,\hat e_{\hat\theta}+\sin\beta\,\hat e_{\hat\phi}\right).
 \label{eq:delta-x}
\end{equation}
Equation~\eqref{eq:F-offworldline} is a local, point-split field and is valid only for small separations. Setting $\Delta\bm x=0$ collapses it back onto $\dd F_0^\mu$ and recovers the local logarithm. Because that singularity is only logarithmic, the mode integrals are continuous at $\delta=0$. The one-sided limits $\delta\to0^\pm$ therefore agree, and we set $\delta=0$ below.

The complete $m$ sum in Eq.~\eqref{eq:addition-theorem} makes the Legendre filter rotationally invariant about the particle.  The $\beta$ average of Eq.~\eqref{eq:F-offworldline} therefore projects onto the part that enters the mode sum and gives $J_0(\kappa a(\psi)r_0\alpha)$, with
\begin{equation}
 a(\psi)=\left[s^2+(1-s^2)\sin^2\psi\right]^{1/2}\,,
 \label{eq:a-psi}
\end{equation}
where $a(\psi)$ is the magnitude of the sphere-tangential projection of $\hat k$.  We now use the asymptotics $P_\ell(\cos\alpha)=J_0(L\alpha)+\Ord(L^{-1})$.  The Bessel orthogonality relation localizes the momentum integral on the shell $\kappa=L/[a(\psi)r_0]$.  Although this calculation formally extends the angular integral over the whole tangent plane, the large-$L$ projection is itself concentrated at $\alpha\sim L^{-1}$.  It therefore probes precisely the region in which the local point-split field is valid.

The ultraviolet measure $\dd\kappa/\kappa$ is scale invariant, so the dependence on the location of the shell, and hence on $a(\psi)$, cancels.  The remaining $\psi$ integral is trivial and gives Eq.~\eqref{eq:Fell-result}.  This derivation fixes the coefficient of the $1/L$ tail; subleading terms depend on the off-worldline extension and are absorbed into the puncture scheme and finite matching coefficient.

\subsection{Puncture and mode-scheme matching}

The sequence $K_{{\rm ct},\ell}$ is not fixed uniquely by the asymptotic $1/L$ tail.  Any sequence with the same large-$\ell$ behavior removes the divergence, while differences at finite $\ell$ shift the finite remainder absorbed into $K_{\rm ren}$.  Fixing the sequence at every $\ell$ amounts to choosing an off-worldline extension of the counterterm. Summing against Legendre polynomials yields the function on the sphere
\begin{equation}
 K_{\rm ct}(\alpha)=\sum_{\ell=0}^\infty K_{{\rm ct},\ell}P_\ell(\cos\alpha),
 \label{eq:puncture-as-synthesis}
\end{equation}
and Eq.~\eqref{eq:K-ct-puncture} follows from the identity
\begin{equation}
 -\ln\sin\frac{\alpha}{2}
 =\frac12+\sum_{\ell\ge1}\frac{L}{\ell(\ell+1)}P_\ell(\cos\alpha).
 \label{eq:lnsin-identity}
\end{equation}
Thus $F^{\hat a}_{\rm punc}(\alpha)=\cD^{\hat a}K_{\rm ct}(\alpha)$ is a field defined on the sphere whose coincidence limit is the local counterterm.  We emphasize that this extension is scheme-dependent. Adding this field before performing the mode sum gives
\begin{equation}
 \sum_{\ell=0}^{\infty}\left(F_\ell^{\hat a}+\cD^{\hat a}K_{{\rm ct},\ell}\right)
 =\lim_{\alpha\to0}\left[\bar{\mathfrak F}^{\hat a}(\alpha)
 +F^{\hat a}_{\rm punc}(\alpha)\right],
 \label{eq:coincidence-collapse}
\end{equation}
where $\bar{\mathfrak F}^{\hat a}$ is the azimuthal average entering Eq.~\eqref{eq:addition-theorem}.

The mode sum has now been traded for the coincidence limit of two known fields.  At finite $\alpha$, the radial integral in Eq.~\eqref{eq:F-offworldline} is $\ln[\Lambda/(a(\psi)r_0\alpha)]$, where $\Lambda=2e^{-\gE}/k_{\min}$.  The sound-speed dependence sits in the finite angular average, for which $\langle\ln(a^2)\rangle_\psi=2\ln[(1+s)/2]$ gives
\begin{equation}
 \bar{\mathfrak F}^{\hat a}(\alpha)\to
 C^{\hat a}\left[\ln\frac{\Lambda}{r_0\alpha}+\ln\frac{2}{1+s}\right].
 \label{eq:psi-average}
\end{equation}
Combining this with $F^{\hat a}_{\rm punc}\to C^{\hat a}(\ln\alpha-\frac12\ln2)$ cancels the local $\ln\alpha$ singularity and produces Eq.~\eqref{eq:collapsed-sum}.

For the straight-line problem, the renormalized force is
\begin{equation}
 F^{\mu,\,\rm flat}_{\rm ren}
 =C^\mu\left[\ln\frac{\Lambda}{r_0}+\ln\frac{\sqrt2}{1+s}\right]
 +\cD^\mu K_{\rm ren}(r_0).
 \label{eq:flat-renormalized}
\end{equation}
The $r_0$ dependence is an artifact of the fiducial tangent-plane construction.  Independence of the physical force from this choice requires
\begin{equation}
 \frac{\dd K_{\rm ren}(\gamma;r_0)}{\dd\ln r_0}
 =\cA(\gamma,v_{\rm rel},c_s)\,.
\end{equation}
This is the RG running of Ref.~\cite{Modrekiladze:2026twz} with the opposite sign because $r_0$ is a position-space factorization scale.

For dust the kinematic support sets $k_\parallel=0$, so the remaining logarithm is an integral over the momentum $k_\perp$ conjugate to the impact parameter.  To compare the hard momentum IR cutoff with the hard impact-parameter IR cutoff of Ref.~\cite{Traykova:2023qyv}, we introduce a simplified version of the force integral, corresponding to the integrated two-body momentum impulse, but with its short-distance behavior smoothed out. The coordinate- and momentum-space integrals then have the asymptotic forms
\begin{align}
 I_b&\propto\int_0^{b_{\rm max}}\dd b\,\frac{b^3}{(b^2+\sigma^2)^2}
 =\ln\frac{b_{\rm max}}{\sigma}-\frac12+\cdots,
 \nonumber\\
 I_k&\propto\int_{k_{\min}\sigma}^{\infty}\dd x\,xK_1^2(x)
 =\ln\frac{2}{k_{\min}\sigma}-\gE-\frac12+\cdots .
 \label{eq:IR-scheme-conversion}
\end{align}
Equating the two representations gives $k_{\min}=2e^{-\gE}/b_{\rm max}$ and therefore $\Lambda=b_{\rm max}$.  This relation is the finite scheme conversion between a hard transverse-momentum cutoff and a hard impact-parameter cutoff. Matching Eqs.~\eqref{eq:flat-renormalized} and \eqref{eq:schematic-matching} now gives Eq.~\eqref{eq:Kren-general-s}.  Equation~\eqref{eq:Kren-shift} follows at $s=0$, $\ell_{\rm UV}=R_c$, and $\cR=\cR_{\rm match}$.

\end{document}

%% file: diagram_library.tex
\usetikzlibrary{arrows.meta,decorations.pathmorphing,shapes.geometric}

\tikzset{
  esf diagram/.style={x=0.80cm,y=0.72cm,baseline=(current bounding box.center)},
  curved graviton/.style={
    draw=black,
    line width=0.52pt,
    double=white,
    double distance=1.35pt,
    decorate,
    decoration={snake,amplitude=1.25pt,segment length=4.8pt}
  },
  causal graviton/.style={
    curved graviton
  },
  causal arrow/.style={
    pos=0.50,
    sloped,
    allow upside down,
    inner sep=0pt
  },
  worldline/.style={draw=black!65,densely dotted,line width=0.62pt},
  source/.style={circle,fill=black,draw=black,inner sep=0pt,minimum size=8.2pt},
  bulk vertex/.style={circle,fill=black,draw=black,inner sep=0pt,minimum size=9.2pt},
  W vertex/.style={circle,fill=white,draw=black,line width=0.68pt,
    inner sep=1.3pt,minimum size=19pt,font=\small},
  response port/.style={circle,fill=white,draw=black,line width=0.65pt,
    inner sep=0pt,minimum size=5.2pt},
  contact vertex/.style={diamond,fill=black,draw=black,inner sep=0pt,minimum size=8pt},
  environment line/.style={draw=black,line width=0.52pt,
    double=white,double distance=1.05pt},
  order label/.style={font=\normalsize},
  small annotation/.style={font=\small}
}

\newif\ifCausalArrowHalo
\CausalArrowHalofalse

\newcommand{\CausalArrowHaloOff}{\CausalArrowHalofalse}

\newcommand{\CausalArrow}{%
  node[causal arrow]{%
    \tikz[baseline=-0.5ex]{%
      \ifCausalArrowHalo
      \draw[
        white,
        line width=0pt,
        -{Stealth[length=8.7pt,width=9.2pt]}
      ]
        (0,0)--(0.01pt,0);
      \fi
      \draw[
        black,
        line width=0pt,
        -{Stealth[length=7pt,width=7.5pt]}
      ]
        (0,0)--(0.01pt,0);
    }%
  }%
}

\newcommand{\ESFOrder}[3]{\node[order label] at (#1,#2) {$#3$};}

\newcommand{\FFourDOneStressMean}{%
\begin{tikzpicture}[esf diagram]
  \node[W vertex] (w1) at (0,0.65) {$W_1$};
  \draw[curved graviton] (-1.05,0.65)--(w1);
  \ESFOrder{0}{-0.45}{\epsilon^{1}q^{0}}
\end{tikzpicture}}

\newcommand{\FNineDOneSKMeanMetric}{%
\begin{tikzpicture}[esf diagram]
  \draw[worldline] (-0.65,0)--(0.65,0);
  \node[source] (sink) at (0,0) {};
  \node[W vertex] (w1) at (0,2.05) {$W_1$};
  \draw[causal graviton] (w1)--(sink) \CausalArrow;
  \ESFOrder{0}{-0.72}{\epsilon^{1}q^{1}}
\end{tikzpicture}}

\newcommand{\FNineDTwoSKResponseChain}{%
\begin{tikzpicture}[esf diagram]
  \draw[worldline] (-0.65,0)--(0.65,0);
  \node[source] (sink) at (0,0) {};
  \node[W vertex] (w2) at (0,1.65) {$W_2$};
  \node[W vertex] (w1) at (0,3.30) {$W_1$};
  \draw[causal graviton] (w1)--(w2) \CausalArrow;
  \draw[causal graviton] (w2)--(sink) \CausalArrow;
  \ESFOrder{0}{-0.72}{\epsilon^{2}q^{1}}
\end{tikzpicture}}

\newcommand{\FTenDOneSKEnvironmentalResponse}{%
\begin{tikzpicture}[esf diagram]
  \draw[worldline] (-0.55,0)--(3.15,0);
  \node[source] (src) at (0,0) {};
  \node[source] (sink) at (2.60,0) {};
  \node[W vertex] (w2) at (1.30,2.05) {$W_2$};
  \draw[causal graviton] (src)..controls(0.35,1.10)and(0.65,1.75)..(w2)
    \CausalArrow;
  \draw[causal graviton] (w2)..controls(1.95,1.75)and(2.25,1.10)..(sink)
    \CausalArrow;
  \ESFOrder{1.30}{-0.72}{\epsilon^{1}q^{2}}
\end{tikzpicture}}

\newcommand{\FTenDTwoSKMeanMetricNonlinearOneSF}{%
\begin{tikzpicture}[esf diagram]
  \draw[worldline] (-0.55,0)--(3.15,0);
  \node[source] (src) at (0,0) {};
  \node[source] (sink) at (2.60,0) {};
  \node[bulk vertex] (v) at (1.30,1.30) {};
  \node[W vertex] (w1) at (1.30,3.00) {$W_1$};
  \draw[causal graviton] (src)..controls(0.35,0.75)and(0.80,1.15)..(v)
    \CausalArrow;
  \draw[causal graviton] (w1)--(v) \CausalArrow;
  \draw[causal graviton] (v)..controls(1.80,1.15)and(2.25,0.75)..(sink)
    \CausalArrow;
  \ESFOrder{1.30}{-0.72}{\epsilon^{1}q^{2}}
\end{tikzpicture}}

\newcommand{\DFDOneWTwoResponse}{%
\begin{tikzpicture}[esf diagram]
  \node[W vertex] (w2) at (0,0) {$W_2$};
  \draw[causal graviton] (-1.20,0)--(w2) \CausalArrow;
  \draw[causal graviton] (w2)--(1.20,0) \CausalArrow;
  \node[small annotation] at (0,-0.90) {deterministic response};
\end{tikzpicture}}

\newcommand{\DFDTwoWTwoContact}{%
\begin{tikzpicture}[esf diagram]
  \node[contact vertex] (c) at (0,0) {};
  \draw[causal graviton] (-1.20,0)--(c) \CausalArrow;
  \draw[causal graviton] (c)--(1.20,0) \CausalArrow;
  \node[small annotation] at (0,-0.90) {contact/Ward term};
\end{tikzpicture}}

\newcommand{\DFDThreeWTwoExchange}{%
\begin{tikzpicture}[esf diagram]
  \node[response port] (p1) at (-0.78,0) {};
  \node[response port] (p2) at ( 0.78,0) {};
  \draw[causal graviton] (-1.78,0)--(p1) \CausalArrow;
  \draw[environment line]
    (p1)--(p2)
    node[pos=0.50,above=5pt,small annotation] {$G_{\rm env}^{R}$}
    \CausalArrow;
  \draw[causal graviton] (p2)--(1.78,0) \CausalArrow;
  \node[small annotation] at (0,-0.90) {excitation exchange};
\end{tikzpicture}}

\newcommand{\FFourDOneSKStressMean}{\FFourDOneStressMean}

\newcommand{\FFourDTwoSKStressSelfGravity}{%
\begin{tikzpicture}[esf diagram]
  \node[W vertex] (w2) at (0,0.65) {$W_2$};
  \node[W vertex] (w1) at (2.05,0.65) {$W_1$};
  \draw[curved graviton] (-1.05,0.65)--(w2);
  \draw[causal graviton] (w1)--(w2) \CausalArrow;
  \ESFOrder{1.02}{-0.45}{\epsilon^{2}q^{0}}
\end{tikzpicture}}

\newcommand{\FFourDThreeSKStressDrivenByParticle}{%
\begin{tikzpicture}[esf diagram]
  \node[W vertex] (w2) at (0,0.65) {$W_2$};
  \node[source] (s) at (2.20,0.65) {};
  \draw[curved graviton] (-1.05,0.65)--(w2);
  \draw[causal graviton] (s)--(w2) \CausalArrow;
  \ESFOrder{1.10}{-0.45}{\epsilon^{1}q^{1}}
\end{tikzpicture}}